\documentclass[conference, a4paper]{IEEEtran}

\usepackage{cite}
\usepackage{amsmath,amssymb,amsfonts}
\usepackage{algorithmic}
\usepackage{graphicx}
\usepackage{textcomp}
\usepackage{xcolor,acronym}

\usepackage{float}
\usepackage{subfig} 

\begin{document}

\acrodef{PPP}[PPP]{Poisson point process}
\acrodef{NPPP}[NPPP]{Non-homogeneous PPP}
\acrodef{PGFL}[PGFL]{probability generating functional}
\acrodef{CDF}[CDF]{cumulative distribution function}
\acrodef{PDF}[PDF]{probability distribution function}
\acrodef{PMF}[PMF]{probability mass function}
\acrodef{PCF}[PCF]{pair Correlation Function}
\acrodef{RV}[RV]{random variable}
\acrodef{SIR}[SIR]{Signal-to-interference ratio}
\acrodef{i.i.d.}[i.i.d.]{independent and identically distributed}
\acrodef{w.r.t.}[w.r.t.]{with respect to}
\acrodef{MAC}[MAC]{Medium Access Control}
\acrodef{BEM}[BEM]{bipartite Euclidean matching}
\acrodef{1D}[1D]{one-dimensional}
\acrodef{2d}[2d]{two-dimensional}
\acrodef{VANET}[VANET]{vehicular ad hoc network}
\acrodef{LT}[LT]{Laplace Transform}
\acrodef{CoV}[CoV]{Coefficient-of-variation}
\acrodef{UE}[UE]{User Equipment}
\acrodef{BS}[BS]{Base Station}
\title{Meta Distribution of SIR in the Internet of Things Modelled as a Euclidean Matching}
  \author{Alexander P. Kartun-Giles, \textit{Member, IEEE}, Konstantinos Koufos, and Sunwoo Kim, \textit{Senior Member, IEEE} %
    \thanks{A. P. Kartun-Giles is with the School of Physical and Mathematical Sciences, NTU Singapore, Singapore, K.~Koufos is with the School of Mathematics, University of Bristol, Bristol, U.K., and S.~Kim is with the Wireless Systems Laboratory, School of Electronic and Electrical Engineering, Hanyang University, Seoul, Republic of South Korea. We thank G. Sicuro and Springer International Publishing AG for allowing reproduction of Fig. \ref{fig:torus} from his PhD thesis \cite{sicuro2017}. This research was supported by the MSIT (Ministry of Science and ICT), Korea, under the ITRC (Information Technology Research Center) support program(IITP-2021-2017-0-01637) supervised by the IITP (Institute for Information \& Communications Technology Planning \& Evaluation). The corresponding author is A.~P.~Kartun-Giles, alexanderkartungiles@gmail.com}}

\maketitle

\begin{abstract}
The Poisson bipolar model considers user-base station pairs distributed at random on a flat domain, similar to matchsticks scattered onto a table. Though this is a simple and tractable setting in which to study dense networks, it doesn't properly characterise the stochastic geometry of user-base station interactions in some dense deployment scenarios, which may involve short and long range links, with some paired very nearby optimally, and others sub-optimally due to local crowding. Since the users will pair one-to-one with base stations, we can consider using the popular bipartite Euclidean matching (BEM) from spatial combinatorics, and study the corresponding (meta) distribution of the signal-to-interference-ratio (SIR). This provides detailed information about the proportion of links in the network meeting a target reliability constraint. We can then observe via comparison the impact of taking into account the variable/correlated short-range distances between the transmitter-receiver pairs on the communication statistics. We illustrate and quantify how the widely-accepted bipolar model fails to capture the network-wide reliability of communication in a typical ultra-dense setting based on a binomial point process. We also show how assuming a Gamma distribution for link distances may be a simple improvement on the bipolar model. Overall, BEMs provide good grounds for understanding more sophisticated pairing features in ultra-dense networks. 
\end{abstract}
\begin{IEEEkeywords}
Matching theory, stochastic geometry, Internet of Things, random geometric graphs, data capacity, interference, wireless networks.
\end{IEEEkeywords}

\section{Introduction}
Geometry plays a fundamental role in the data capacity of dense wireless networks by requiring the excessive division of available spectrum into multiple adjacent, non-overlapping channels~\cite{ge2016}. The notion of \textit{area spectral efficiency} (ASE) has been introduced to measure the {\textit{bits per second per Hz per unit volume}} the network can move from source to destination. There is an upper bound representing the fundamental limit of densification, below which ASE will improve by simply adding, at financial cost, more base stations~\cite{franceschetti2009,alammouri2017}. 

In this setting, stochastic geometry yields simple but insightful expressions for the ASE, rate distribution and coverage probability of wireless communication networks~\cite{alammouri2017, gupta2000,lu2015,haenggi2016,haenggi2012,gupta2015, wang2018,knight2017,giles2016, giles2015,kartungiles2019,koufos2016,koufos2018,koufos2019,koufos20192, ganti2011}. Traditional spatial models, e.g., the Poisson bipolar model, take into account the effect of interference on the typical link, but consider only trivial notions of its length. Put simply, the base stations in the bipolar model are distributed at random, and users are then generated at a fixed distance away from each, and at a random angle. For example, this distance is either fixed and known~\cite{haenggi2016}, or is variable according to some standard distribution~\cite{weber2012, weber2005}. Nevertheless, perturbed resource allocation can  change the pair associations, and this may have a knock-on impact throughout the network. This may lead to very sophisticated random link distance distributions, and therefore to a sophisticated meta distribution of the SIR~\cite{haenggi2016}. It is important to develop research in this area along the lines of more arbitrary pairing ideas between users and base stations which capture the geometric confinement of ultra-dense communication, which leads to better understanding of this scenario generally.

\begin{figure}[!t]
  \centering
       \includegraphics[width=\columnwidth]{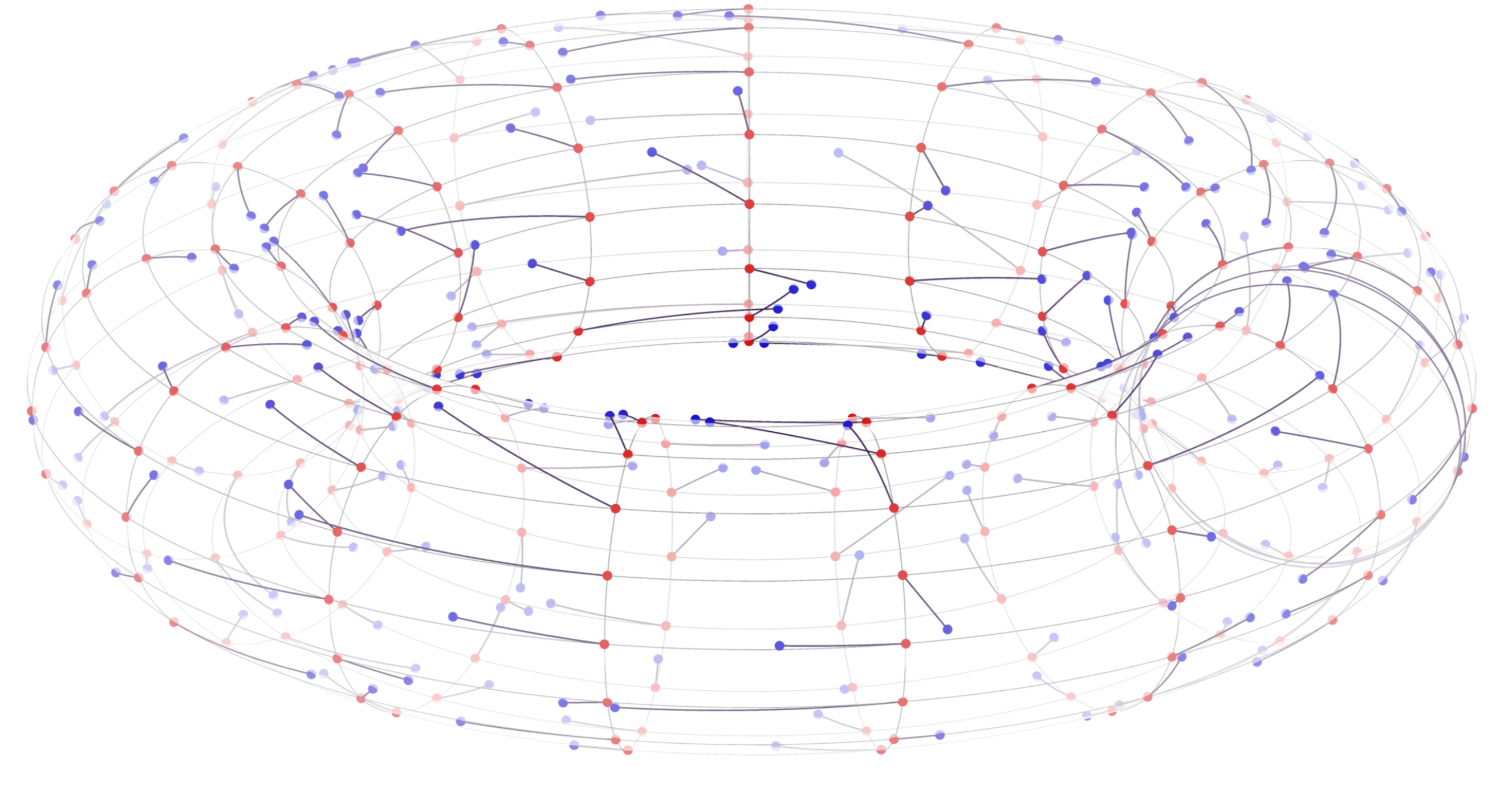}
          \caption{An example realisation of a bipartite Euclidean matching of $N=225$ pairs of points, distributed uniformly at random, from a recent work on the Euclidean matching problem \cite{sicuro2017}. The sum of the interpoint distances is near minimal over the set of all matchings between the points.}
\label{fig:torus}
\end{figure}

In this paper, and following on from previous work \cite{kartungiles2018,kreacic2019}, we consider \acp{UE} and \acp{BS} modeled by two binomial point processes in the \ac{2d} Euclidean space without boundary (this will look like a Poisson process at sufficient density). The two processes are paired-off in an edge-independent way, i.e., no edge is incident to more than one point, known as a \ac{BEM}, shown in Fig. \ref{fig:torus}. We pair the point processes to minimize the aggregate pair distances in this setting. This leads to significant ``ground state entropy'', where more than one good configuration is possible (though only one absolutely optimal, due to the continuous link lengths) ~\cite{villani2003}. Since a \ac{BS} can be the nearest \ac{BS} for more than one \acp{UE}, the link distances of nearby communication pairs in this setting become correlated.

The main argument of this article is that the BEM incorporates both variable link distances and Euclidean correlations. These are crucial aspects of engineering geometrical configurations, making this both novel and significant. It also resembles very mathematical deep ideas in condensed matter physics, with the hope of shedding light on the depth of this sort of communication problem~\cite{mezard1988}. This makes it much more realistic than other variable distance models, e.g., based on the Rayleigh distribution~\cite{weber2012, weber2005}. Forming links in a dense network according to a \ac{BEM} also agrees fairly enough with intuition, i.e., it is natural to assume that every node will try to pair with a nearby available, i.e., unpaired node. Therefore, the distribution of link distances in an ultra-dense ad hoc network shall be better captured by a \ac{BEM} rather than the traditional bipolar or variable link distance models. This has been studied in two other articles, first by Kartun-Giles, Kim and Jayaprakasam \cite{kartungiles2018}, and later by Kreacic and Bianconi \cite{kreacic2019}. We attempt to go some way to introducing interference as a key idea, which is omitted in those works due to tractability concerns, and using the popular meta distribution framework of Haenggi \cite{haenggi2016}.

The simplistic bipolar model assumes an equal and known distance for all communication pairs for the sake of tractability of, e.g., the statistics of the meta distribution. The study in~\cite[Eq.~(3.29)]{haenggi2009} points out that the nearest-neighbor distance, which scales as $\lambda^{-1/2}$ in the plane, where $\lambda$ stands for the point process intensity, could be integrated into the calculation of the outage probability. Later, the study in~\cite[Section 4.2]{weber2012} uses the contact distribution of planar \ac{PPP}, i.e., the Rayleigh distribution, and compares the outage probability with and without variable link distances as $\lambda\rightarrow 0$. Both studies~\cite{haenggi2009, weber2012}, however, neglect the inherent correlations of link distances in transmitter-receiver (Tx-Rx) pairing in ad hoc networks. This has subtle but, as we will show, very important effects on the statistics of the meta distribution. 
%


BEMs are relevant in both machine-to-machine communications and wireless sensor networks, where a massive number of spatially embedded devices distribute measurement information over their network. An upper bound on the maximum data rate concerns an effective ``transport plan'', which is optimized given the constraints each interferer places on its neighboring transmitters. Certain transport plans will lead to some Tx-Rx pairs failing to meet an SIR with given reliability. The meta distribution of the SIR is thus an important idea to introduce in this setting. See also the related \textit{spatial outage capacity} studies~\cite{kalamkar2018,kalamkar2019}. The integration of \emph{variable and correlated} link distances with fading and interference into the theory of the meta distribution of the SIR is precisely the contribution of our work. This constitutes a preliminary but important step towards the understanding of the impact of more realistic spatial models into the performance of wireless ultra-dense networks. Preliminaries are discussed in Section II. Meta distribution and spatial outage capacity models for BEMs are investigated in Section~\ref{sec:model}. Section IV concludes this paper. 

\section{Preliminaries} 
\label{sec:Distances}
With intensity $\lambda > 0$ and $N\sim \text{Binomial}(\lambda)$, consider the Binomial point process (BPP) $\mathcal{X}_{N} \subset [0,1]^{d}$ of $N$ points, and the BPP $\mathcal{Y}_{N} \subset [0,1]^{d}$, also of $N$ points, drawn from the hypercube with periodic boundary conditions, which is equivalent to a torus when $d=2$, with $2N$ points distributed uniformly at random over its (flat) surface. The sets $\mathcal{X}_N$ and $\mathcal{Y}_N$ are independent of each other. Form a \textit{perfect bipartite matching} $\mathcal{M}_{N}$ by assigning $N$ of the pairs with one end in each of $\mathcal{X}_N$ and $\mathcal{Y}_N$ in such a way that every point is incident to exactly one pair.


For now, ignore the impact of fading and interference. Denote the Euclidean lengths of the edges in a BEM by $d_{1},d_{2},\dots,d_{N}$. For each matching, we therefore have a total length $L_{\mathcal{M}} = \sum_{i}d_{i}$ and a one-hop capacity \begin{equation}C_{\mathcal{M}} = \sum_{i=1}^{N}\log_{2}(1+d_{i}^{-\eta}),\end{equation} where $\eta$ is the propagation pathloss exponent. In the expression of $C_{\mathcal{M}}$, various constants like the bandwidth, the noise and the transmit power levels, for simplicity, have been omitted. The perfect matching which minimises $L_{\mathcal{M}}$ does not necessarily maximize $C_{\mathcal{M}}$. However, since the function $\log_{2}(1+d_{i}^{-\eta})$ is monotonically decreasing and convex in $d_i$ for positive $\left\{d_i,\eta\right\}$, the Jensen's inequality yields 
\begin{equation}
  \label{eq:Cm}
C_{\mathcal{M}} \geq N \log_2\left(1+\left(\frac{L_{\mathcal{M}}}{N} \right)^{-\eta} \right). 
\end{equation}

Eq.~\eqref{eq:Cm} shows that minimizing the sum-distances $L_{\mathcal{M}}$ maximizes a lower bound of the one-hop sum-capacity, with equality observed for equal $d_i \forall i$, i.e., the bipolar model. Despite its importance from a theoretical point of view, this analysis, see~\cite{kartungiles2018} for further details, is rather trivial. It neglects the impact of randomness in the fading channel, and the impact of interference. Also, the lower bound attained by Jensen's inequality might not be tight. Nevertheless, its outcome agrees with intuition: Pairing the communication nodes in a network such that the parameter $L_{\mathcal{M}}$ is minimized should be beneficial for the aggregate one-hop sum-capacity $C_{\mathcal{M}}$ too. 



Next, we turn our attention from the aggregate capacity to the rate distribution across the network. Mathematically, the \textit{link reliability} is defined as the conditional probability of successful decoding  $P_s\left(\theta\right)=\mathbb{P}\left({\text{SIR}}>\theta|\Phi\right)$, where $\theta$ is the operation threshold at the receiver, and  $\Phi=\mathcal{X}_N\cup\mathcal{Y}_N$, in our case, corresponds to the point process of transmitters and receivers in a \ac{BEM}. Assuming unit-mean exponential random variables $h_k$ for the channel power fading and Bernoulli variates $\xi_k$ with parameter $\xi$ for the activity of each interferer, the reliability of a link in a BEM can be read as~\cite[Appendix]{haenggi2016}  %
\begin{equation}
  \label{eq:Ps0}
P_s\left(\theta\right)= \mathbb{P}\left( \frac{h r^{-\eta}}{\sum\limits_{x_k\in\Phi \backslash\{o\}} h_k \xi_k x_k^{-\eta}} \geq \theta | \Phi\right),
\end{equation}
where  $h$ is the channel power fading following a unit-mean exponential distribution and $r$ is the distance of the Tx-Rx link whose reliability is computed. Additionally, from the point process $\Phi$ we have excluded the Tx of the link which is assumed to be located, without any loss of generality, at the origin $o$. Finally, $x_k$ are the locations of interferers conditioned on the realization of $\Phi$.

After averaging over the activity and fading distributions of the interferers, Eq.~\eqref{eq:Ps0} yields 
\begin{equation}
  \label{eq:Ps}
  \begin{array}{ccl}
    P_s\left(\theta\right) &=&  \mathbb{E}_{\xi_k, h_k}\left\{\exp\left( -\theta r^\eta \sum\limits_{x_k\in\Phi \backslash\{o\}} h_k \xi_k x_k^{-\eta} \right) \right\}\\
   &=& \mathbb{E}_{\xi_k, h_k}\left\{ \prod\limits_{x_k\in \Phi \backslash\{o\}} \exp\left( -\theta r^\eta h_k \xi_k x_k^{-\eta} \right) \right\}\\
    &=&  \mathbb{E}_{h_k}\left\{ \prod\limits_{x_k\in\Phi \backslash\{o\}}\left(1-\xi + \xi \exp\left( -\theta r^\eta h_k x_k^{-\eta} \right)\right) \right\} \\
     &=& \prod\limits_{x_k\in\Phi \backslash\{o\}}\left(1-\xi + \frac{\xi}{1+\theta r^\eta x_k^{-\eta}} \right).
    \end{array}
\end{equation}

The complementary \ac{CDF} of the \ac{RV} $P_s\left(\theta\right)$ in Eq.~\eqref{eq:Ps}, i.e., $\mathbb{P}\left(P_s\left(\theta\right)>u\right), u\in\left[0,1\right]$ is usually referred to as the \textit{meta distribution} of the SIR. Given the operation threshold $\theta$, it represents the percentage of random spatial realizations where the Rx of the considered link meets the SIR target $\theta$ with probability  larger than $u\in\left[0,1\right]$. The probability is calculated over the distributions of fading and activity as in Eq.~\eqref{eq:Ps}. For ergodic point processes, this percentage is also equal to the fraction of links achieving a reliability higher than $u$ for each spatial realization, i.e., for fixed but unknown locations of Tx and Rx in the BEM. 

The moments of the meta distribution of the SIR, $M_b\left(\theta\right)= \mathbb{E}\{P_s\left(\theta\right)^b\}$, have been investigated for Poisson bipolar and heterogeneous wireless networks using the \ac{PGFL} of \ac{PPP}~\cite{haenggi2016,haenggi2012,kalamkar2018}. Assuming that the locations of interferers follow a \ac{PPP}, which is a reasonable assumption in the ultra-dense limit $N\to\infty$, the moments of the meta distribution can be calculated by raising Eq.~\eqref{eq:Ps} to the $b$-th power and expressing the spatial average using the \ac{PGFL} of the \ac{PPP}~\cite{haenggi2016}.

\begin{align} \label{eq:Mb}
      M_b\left(\theta\right) =& \mathbb{E}_\Phi \left\{ \prod\limits_{x_k\in \Phi \backslash\{o\}} \left(1-\xi + \frac{\xi}{1+\theta r^\eta x_k^{-\eta}} \right)^b\right\}\\
      =&  \exp\left.\Bigg(-\lambda \right. \nonumber \\ & \left. \times  \int\limits_0^{2\pi}\int\limits_0^\infty1-\left(1 - \xi + \frac{\xi}{1+\theta r^\eta x_k^{-\eta}} \right)^b\right.  \left. x{\rm d}x {\rm d}\phi \right.\Bigg)  \\ 
       =& \exp\left(-2\lambda\pi \int\limits_0^\infty\left(1-\left(1 - \frac{\xi \theta r^\eta }{x^\eta+\theta r^\eta}\right)^b\right) x{\rm d}x \right),
\end{align}
where $\lambda$ is the intensity of interferers. The above equation indicates that the statistics of the random link distance $r$ are therefore central. 

\section{Meta Distribution for \acp{BEM}}
\label{sec:model}
We consider a perfect \ac{BEM} $\mathcal{M}_{N}$ between the transmitters and receivers minimizing the sum of Euclidean distances $L_{\mathcal{M}} = \sum_{i}d_{i}$, and we will seek the proportion of links in the matching which are able to achieve an SIR greater than $\theta$ with probability at least $u$. Let us assume a realization of $\mathcal{X}_N$ and $\mathcal{Y}_N$ modeling the set of transmitters and receivers respectively in a wireless ad hoc network over the unit square $\left[0,1\right]^2$. Given $N$, we deploy uniformly at random within the unit square $N$ transmitters, $\left(N-1\right)$ receivers and we also add an extra receiver at the origin $o\equiv(1/2,1/2)$. Under periodic boundary conditions, all receivers experience the same interference field, after averaging over the point processes $\mathcal{X}_N$ and $\mathcal{Y}_N$. In addition, we assume that the process of line segments formed by the Tx-Rx communication pairs in a BEM is ergodic. This is not as straightforward to prove as in Poisson bipolar or cellular networks~\cite{ganti2010, haenggi2016}, because nearby links span correlated distances in a \ac{BEM}. Under the ergodicity assumption, the receiver at the origin is hereafter referred to as the \textit{typical receiver}, and the meta distribution of the SIR at the \textit{typical receiver} will correspond to the fraction of links in the BEM experiencing certain reliability.  

The bipolar model assumes a fixed and known link distance $r=R$. Then, the moments of the meta distribution of the SIR follow from Eq.~\eqref{eq:Mb} after substituting $r=R$ and using standard steps to calculate the integral therein~\cite{ganti2010, haenggi2016}. Finally, we get
\begin{equation}
\label{eq:MetaBipolar}
M_{b}(\theta) = \exp(-C_b(\theta) R^2), 
\end{equation}
where \begin{equation}C_b(\theta) = \frac{\lambda\pi \Gamma(\delta)\Gamma(1-\delta)}{\theta^{-\delta} (\delta b \xi)^{-1}} {}_2F_1(1-b,1-\delta,2;\xi)\end{equation} and   $\delta=2/\eta$. Note that the above equation is equivalent to~\cite[Eq.~(5)]{haenggi2016} using the diversity polynomials.

The link distance $R$ in the traditional bipolar model is assigned a value either independent of the intensity $\lambda$, or proportional to the nearest-neighbor distance $R\propto\lambda^{-1/2}$~\cite[Eq.~(3.29)]{haenggi2009}. The latter agrees with intuition, i.e., a higher intensity of interferers should come along with shorter distances for the Tx-Rx link. Since $R^2\propto\lambda^{-1}$, we see from Eq.~\eqref{eq:MetaBipolar} that the average success probability for nearest-neighbor communication becomes independent of $\lambda$. Moreover, the scaling $R\propto\lambda^{-1/2}$ ignores the correlations of link distances inherent in a \ac{BEM}. 

Searching for more accurate link distance models, we note that in the ultra-dense limit $N \rightarrow \infty$, the sum of link distances in \acp{BEM} scales as $\sqrt{\lambda \log\lambda/2\pi}$~\cite{ajtai1984,caracciolo2014} and~\cite[Eq.~(6)]{caracciolo2015}. Therefore, the mean link distance scales as $\sqrt{\log\lambda/2\pi\lambda}$. One may substitute this value instead of $R$ in~\eqref{eq:MetaBipolar} and proceed with the calculation of the moments. For instance, the first moment takes the following simple form: 
\begin{equation}
\label{eq:SuccessBipolar}
M_1(\theta) = \frac{1}{\lambda^c},  \lambda\rightarrow\infty,  c=\frac{\pi\xi\theta^\delta}{\eta}\csc\frac{2\pi}{\eta}. 
\end{equation}

Eq.~\eqref{eq:SuccessBipolar} can be seen as a complementary result to~\cite[Eq.~(3.29)]{haenggi2009}. It indicates that the average  probability of successful decoding decreases for denser bipolar networks, while all other parameters remain fixed. The new bipolar model still  assumes a fixed and known link distance, but unlike the traditional one, it takes into account some of the correlation of link distances.

The new bipolar model in Eq.~\eqref{eq:SuccessBipolar} offers an improved estimate for the link distance in a \ac{BEM}, but it still neglects the fact that the link distances would naturally follow a distribution. Inspired by the void probability of planar \ac{PPP}, one popular link distance model in wireless communications research is the Rayleigh distribution. In our system setup, we will consider the Rayleigh \ac{PDF} $f(r)$ with mean equal to $\sqrt{\log\left(\lambda\right)/2\pi\lambda}$: 
\begin{equation}
  f\left(r\right)=\frac{\pi^2}{\log\lambda}\lambda r  \exp{\left(-\frac{\pi^2}{2\log\lambda}\lambda r^2\right)}.
\end{equation}
\begin{figure}[!t]
  \centering
    \includegraphics[width=3.7in]{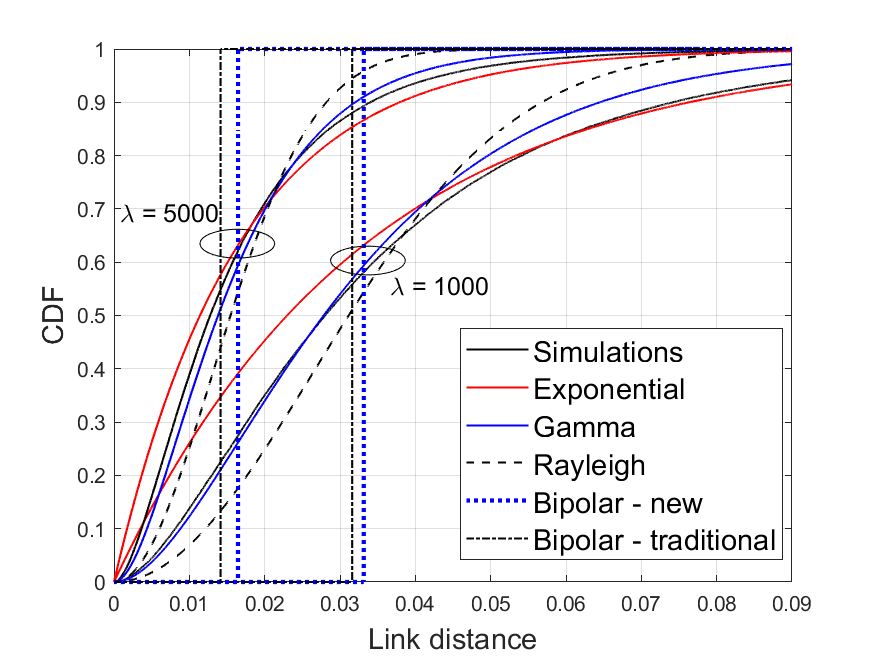}
    \caption{Approximating the link distance distribution in BEMs using various models. We use the toolbox 'matchpairs' in MatLab (R2019b) which implements an algorithm of Duff and Koster \cite{duff2001} to generate the Tx-Rx pairs (in the unit square) conditioned on the realization of the point processes $\Phi=\mathcal{X}\cup \mathcal{Y}$. For the traditional bipolar model the link distance equals $\lambda^{-1/2}$, and for the new bipolar model the link distance is $R=\sqrt{\log(\lambda)/2\pi\lambda}$. Since the realizations of the BPPs are generated within the unit square, we have $N=\lambda$.} 
\label{fig:Distances}
\end{figure}
\begin{figure*}[!t]
  \centering
  \subfloat[$\lambda=500, \xi=0.5, \theta=1$] {\includegraphics[width=2.in]{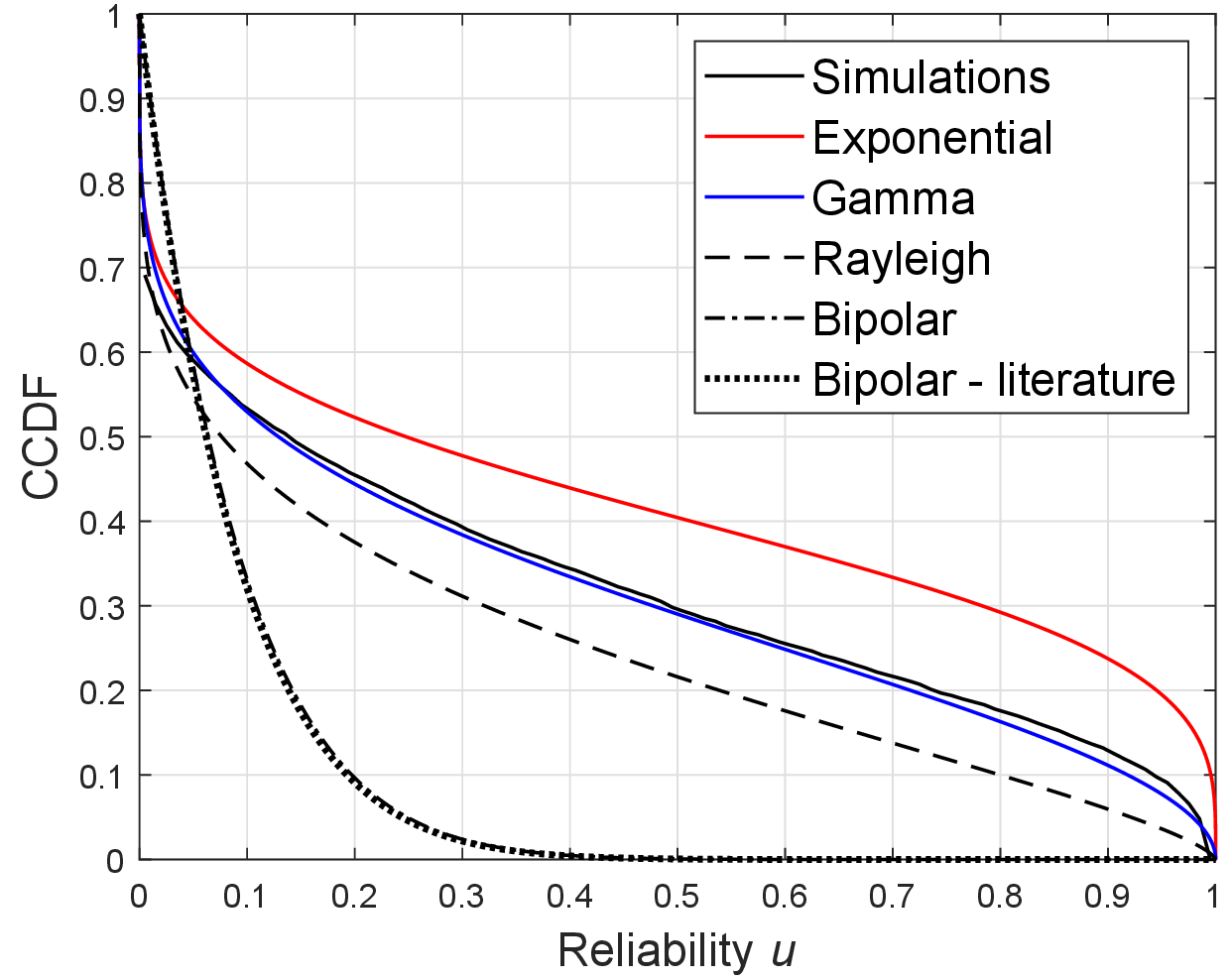} \label{fig:MetaDistributions}} \hfil
  \subfloat[$\lambda=1000, \xi=0.25, \theta=1$] {\includegraphics[width=2.2in]{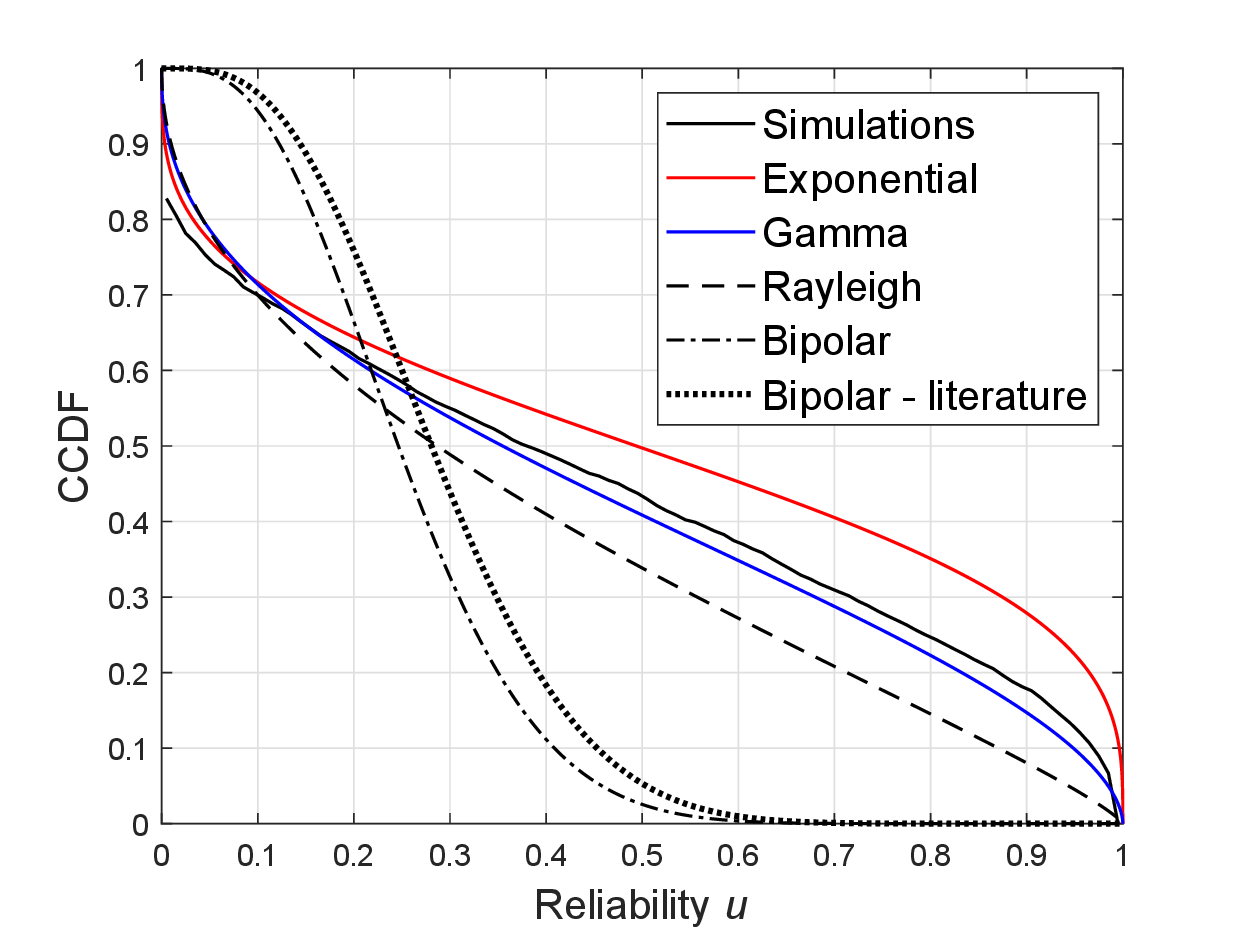} \label{fig:MetaDistributions2}} \hfil
    \subfloat[$\lambda=1000, \xi=0.25, \theta=0.1$] {\includegraphics[width=2.25in]{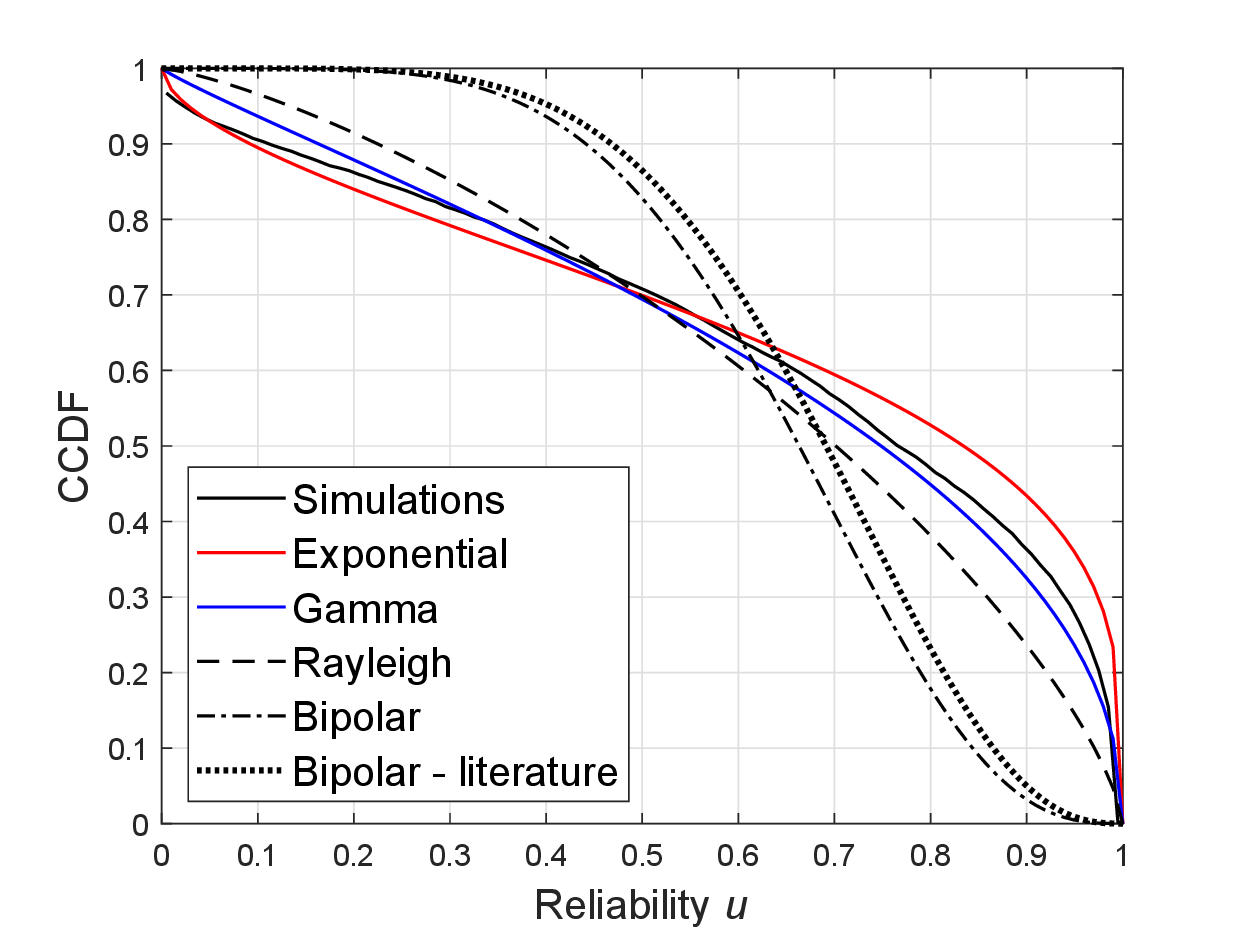} \label{fig:MetaDistributions3}}
 \caption{The meta distribution of the SIR in BEMs within the unit square using simulations and approximations based on various link distance models. $10000$ random spatial configurations are generated. For each configuration, we simulate the probability of successful reception (or reliability) at the typical receiver over $1000$ independent activity and fading realizations. The transmitter associated to the typical receiver is always active. The complementary \ac{CDF} of the histogram of the reliability is plotted with bin size $0.01$. $\eta=4$. For more details, see also the caption of Fig.~2.}
\end{figure*}

The moments of the meta distribution for Rayleigh distributed link distances are calculated after substituting $r$ instead of $R$ in Eq.~\eqref{eq:MetaBipolar} and averaging over the $f(r)$ above, yielding 
\begin{equation}
\label{eq:MetaRayleigh}
M_b\left(\theta\right) = \frac{\pi^2\lambda}{\pi^2\lambda+2\log\left(\lambda\right) C_b\left(\theta\right)}.
\end{equation}

In ultra-dense \acp{BEM}, the link distances scale linearly near the origin, i.e., $f(r) \sim r$ as $r \to 0, \lambda\to\infty$~\cite[Eq.~(7)]{mezard1988}. Therefore, the Gamma distribution \begin{equation}f\left(r\right)=\frac{x^{\nu-1} e^{-x/\beta}}{\beta^\nu \Gamma\left(\nu\right)}\end{equation} with scale $\nu=2$ might be a good candidate model for the distribution of link distances. The shape $\beta$ can be set to match the mean value of link distances in a BEM, i.e., $\beta=\sqrt{\log(\lambda)/8\pi\lambda}$. For the Gamma distribution model, we average Eq.~\eqref{eq:MetaBipolar} over the distribution $f\left(r\right)=r \exp\left(-r/\beta\right)/\beta^2$, yielding
\begin{equation}
  \label{eq:MetaGamma}
 M_b\left(\theta\right) =  \frac{1}{2 \beta^2 C_b} - \frac{\sqrt{\pi}}{4\beta^3 C_b^{3/2}} \exp\left(\frac{1}{4C_b\beta^2}\right) {\text{erfc}}\left(\frac{1}{2\beta\sqrt{C_b}}\right).
\end{equation}

In Fig.~\ref{fig:Distances}, we see that the Gamma distribution provides better fit to the simulated link distance distribution than the Rayleigh distribution. For completeness, we have also included in Fig.~\ref{fig:Distances} the link distance distribution models associated with the exponential distribution, and the unit-step functions pertinent to the two bipolar models. For the exponential link distance model, the moments of the meta distribution are calculated by substituting $r$ instead of $R$ in Eq.~\eqref{eq:MetaBipolar} and averaging over an exponential distribution $f\left(r\right)=\tau \exp(-\tau r)$ with rate $\tau=\sqrt{2\pi\lambda/\log(\lambda)}$. Thus, 
\begin{equation}
\label{eq:MetaExponential}
M_b\left(\theta\right) = \exp\left({\frac{\tau^2}{4C_b}}\right) {\text{erfc}}\left( \frac{\tau}{2 \sqrt{C_b}}\right) \sqrt{\frac{\pi \tau^2}{4C_b}},
\end{equation}
where $C_b\left(\theta\right)\equiv C_b$ for brevity and \begin{equation}{\text{erfc}}\left(z\right)=\frac{2}{\sqrt{\pi}}\int\nolimits_z^ \infty e^{-t^2}{\rm d}t\end{equation} is the complementary error function. 

In Fig.~\ref{fig:MetaDistributions} we simulate the meta distribution of the SIR in a BEM and approximate it using the various link distance models presented above. To generate the approximation for each model, we calculate the first two moments of the meta distribution and use them to fit a Beta distribution, as suggested by Haenggi in~\cite{haenggi2016}. We have checked that the Beta distribution provides a very good fit to the simulated meta distribution for all considered models. Note that due to the simplicity of Eq.~\eqref{eq:MetaRayleigh}-\eqref{eq:MetaExponential}, higher moments of the meta distribution can be easily computed and used to numerically invert its characteristic function~\cite[Eq:~(12)]{haenggi2016}. We see in Fig.~\ref{fig:MetaDistributions} that the performance of the bipolar models is poor over the full range of the distribution, and therefore are clearly unsuitable to model the SIR in \acp{BEM}. The other link distance models can all capture the trend of the meta distribution, with the Gamma model being almost a perfect fit, given we are not using a more advanced technique to capture also the variance of the link distance distribution. 

Let us assume that the network intensity $\lambda$ increases while the product $\lambda\xi$ and subsequently the first moment of the meta distribution $C_1(\theta)$ is kept fixed. For a bipolar model with link distance $R$ independent of the intensity as in Eq.~(4), the average success probability should not change. On the contrary, in \acp{BEM} the average success probability increases for denser networks. This is because the link distance, on average, decreases, while the mean interference level remains the same, compare  Fig.~\ref{fig:MetaDistributions} with Fig.~\ref{fig:MetaDistributions2}. Another remark is that denser networks augment the difference between the traditional and the new bipolar model. Finally, in Fig.~\ref{fig:MetaDistributions3}, we depict the meta distributions for a lower operation threshold $\theta$ illustrating that the performance predictions of the Gamma distribution model remain good.

The promising results obtained so far have motivated us to further look into the properties of the meta distribution in a BEM using the Gamma distribution model for the link distances and Eq.~\eqref{eq:MetaGamma}. To give an example, the behavior of the meta distribution $M_b(\theta)$ as $u\to 1$ sheds some light on the fraction of links that can maintain ultra-reliable connectivity for operation threshold $\theta$. This behavior might for instance be useful while determining medium access control parameters in the network such as the activity probability in Aloha. Following similar steps used in~\cite[Appendix~C]{kalamkar2018}, we have obtained the following approximation for ultra-reliable connectivity in BEMs: 
\begin{equation}
  \label{eq:ultradense}
M_b(\theta) \sim \frac{4 (1-u)^\delta}{\log(\lambda) \Gamma(1-\delta)\Gamma(1+\delta)(\theta\xi)^\delta }, u\to 1.
  \end{equation}

The above equation shows that the fraction of successful communication links in a BEM reduces logarithmically with the intensity $\lambda$, while in~\cite[Eq.~(43)]{kalamkar2018} it is found to be independent of $\lambda$ for an equal intensity of transmitters and receivers. This is due to the logarithmic correction of the mean link distance in BEMs~\cite[Eq.~(6)]{caracciolo2015}.

\begin{figure}[!t]
  \centering
    \includegraphics[width=3.7in]{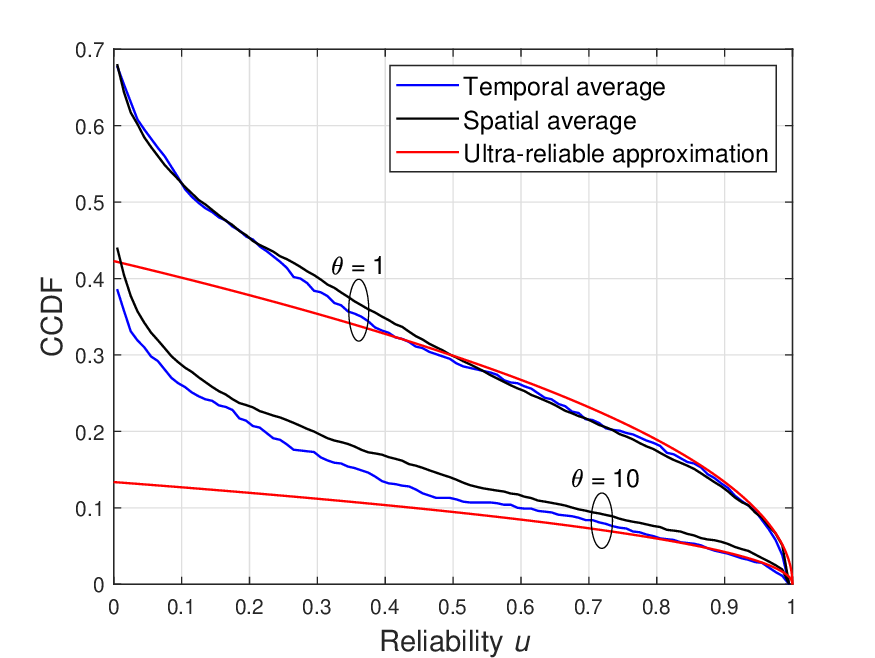}
    \caption{Comparing the meta distribution of the SIR over time (temporal average at the origin) and space (spatial average across the unit square), and also testing the accuracy of the ultra-dense approximation in Eq.~\eqref{eq:ultradense} for $\lambda=5000, \xi=0.5$ and $\eta=4$.}
\label{fig:MatchingsErgodicity}
\end{figure}

Before concluding, it is illustrated in Fig.~\ref{fig:MatchingsErgodicity} that the temporal and spatial distributions of the SIR are close to each other, providing at least some numerical evidence about ergodicity in our system model setup. In the same figure, the quality of the approximation in Eq.~\eqref{eq:ultradense} is also tested, and it is found to perform remarkably well in the ultra-reliable regime $u\to 1$. Despite the various approximations involved in the derivation of Eq.~\eqref{eq:ultradense}, this simplified expression could be of use in the design of wireless ad hoc networks modeled by BEMs, e.g., it can be incorporated into the performance evaluation framework of the spatial outage capacity~\cite{kalamkar2018,kalamkar2019}.  

\section{Conclusions}\label{sec:conclusion}
Two random sets of devices embedded in the Euclidean plane are paired according to a bipartite Euclidean matching (BEM), and the meta distribution of the SIR is studied. We ask what proportion of links have a probability of reliable communication under random activity and Rayleigh fading. The best (i.e. shortest) matching introduces correlated link distances, which, unlike the bipolar model, are all unique. The traditional link distance distribution models for ad hoc networks, i.e., the Rayleigh and the bipolar models, are insufficient approximations in this setting. Moreover, the geometrical configuration of the nodes and their effect on the communication properties of the best matching is highly sophisticated, leading to interesting research questions linking matching theory and the internet of things. We demonstrate this with analysis and simulations, using the Gamma distribution as a temporary test model of the typical link distance. The parameters of the Gamma distribution are fitted to match the geometric properties of ultra-dense BEMs. The variance of link distances in \acp{BEM} is an important open problem, allowing us to build better models for the meta distribution for all intensities. This still only considers links as independent random variables, so it is important to involve mathcing theory mor directly, which would be a key link between statistical physics, and the future internet of things. We hope that this paper opens a bridge between complexity science, and communication theory, via the fascinating Euclidean matching problem which has huge importance for the theory of future geometrically confined networking technologies.

\end{document}